\documentclass[a4paper,11pt]{article}
\usepackage{pos}
\usepackage{multirow}
\usepackage{float}
\usepackage{arydshln}
\usepackage{colortbl}
\usepackage{amsmath}
\usepackage{dsfont} 
\usepackage{hyperref}
\usepackage{graphicx}
\usepackage{enumitem}
\usepackage{arydshln}
\usepackage{mathtools}
\usepackage{bbold}
\usepackage{soul}
\usepackage{slashed}
\usepackage{amsmath,bm}
\usepackage{yfonts}
\usepackage{tikz}
\usepackage{tikz-feynman}
\usepackage{lipsum}
\usepackage{slashed}
\usepackage{mathtools}
\usepackage{comment}
\usepackage{physics}

\newcommand{\be} {\begin{equation}}
\newcommand{\ee} {\end{equation}}
\newcommand{\bea} {\begin{eqnarray}}
\newcommand{\eea} {\end{eqnarray}}

\newcommand{\cR}{{\mathcal R}}

\newcommand{\cL}{{\mathcal L}}

\newcommand{\cA}{{\mathcal A}}
 
\newcommand{\cM}{{\mathcal M}} 
\newcommand{\llpair}{\bar\ell\ell}
\newcommand{\mmpair}{\bar\mu\mu}

\numberwithin{equation}{section}
\numberwithin{figure}{section}
\numberwithin{table}{section}

\newcommand{\cQ}{{\mathcal Q}} 

\newcommand{\cN}{{\mathcal N}}

\title{Short- vs Long-Distance Dynamics in $b \rightarrow s\bar{\ell} \ell$ Decays}
\author*[a]{Arianna Tinari}
\affiliation[a]{Physik-Institut, Universit\"at Zu\"rich, CH-8057 Z\"urich, Switzerland}
\emailAdd{arianna.tinari@physik.uzh.ch}

\abstract{The tension of $B\to K^{(*)}\bar\ell\ell$ decays with the Standard Model (SM) can be attributed to a short-distance (SD) $b s\bar\ell\ell$ interaction.
We show two methods to disentangle this effect from long-distance (LD) dynamics. Firstly, we perform a comparison of the inclusive $b\to s\bar\ell\ell$ rate at high $q^2=m^2_{\ell\ell}\geq 15~\rm GeV^2$ with a determination based on data on the leading exclusive modes, finding a $\sim 2\sigma$ discrepancy. Secondly, we analyze the exclusive $B\to K^{(*)}\bar\ell\ell$ spectrum in the entire $q^2$ region. With a dispersive parametrization of the charmonia resonances, we extract the non-SM contribution to the Wilson coefficient $C_9$ for every bin in $q^2$. The result is compatible with the SD hypothesis and the inclusive determination. Finally, with the aim of having better control over LD effects that mimic the $C_9$ contribution, we estimate the size of a few charm-rescattering processes in $B\to K\bar\ell\ell$.}

\FullConference{42nd International Conference on High Energy Physics (ICHEP2024)\\
18-24 July 2024\\
Prague, Czech Republic\\}

\begin{document}
\maketitle

\section{Introduction}

Flavor-changing neutral transitions such as $b\to s\llpair$ are prime candidates in the search for Beyond the Standard Model (BSM) Physics since they show enhanced sensitivity to short-distance (SD) physics. 
Precise SM tests are confined to  $q^2 \lesssim 6-8\,$GeV$^2$ (low-$q^2$ region) and $q^2 \gtrsim 14-15\,$GeV$^2$ (high-$q^2$ region), where $q^2$ is the invariant mass of the lepton pair, because of the presence of narrow charmonium resonances in the central region.  

On the experimental side, in the last few years measurements of rates and angular distributions of the exclusive $B \to K^{(*)}\mmpair$ decays by LHCb~\cite{LHCb:2013ghj,LHCb:2014cxe,LHCb:2015svh} have shown tensions with the SM, especially in the low-$q^2$ region. At the inclusive level, a tension in the rate at high $q^2$ has been observed in \cite{Isidori:2023unk}.

On the theoretical side, the challenge lies in the difficulty of estimating the non-perturbative contributions, i.e.~the local form factors and the non-local matrix elements of a certain class of operators, the four-quark operators. In order to distinguish a possible SD contribution from the long-distance (LD) effects, it is helpful to look at complementary observables at the exclusive or inclusive level and at low or high $q^2$, as they show different sensitivity to SD versus LD physics, and are affected by different uncertainties.  

We first look at the inclusive decay rate, $\Gamma(B\to X_s \llpair)$, in the high-$q^2$ region; this is usually treated with an operator product expansion (OPE) in $1/m_b$, but in the high-$q^2$ region, the rate is affected by large hadronic uncertainties as it is very sensitive to power corrections. 
It has been shown in \cite{Ligeti:2007sn} that normalizing $\Gamma(B\to X_s \bar{\ell} \ell) $ to $\Gamma(B \to X_u \ell \bar{\nu})$ significantly reduces the uncertainties.

Secondly, we look at the exclusive level $B \to K^{(*)} \llpair$ in the entire $q^2$ spectrum \cite{Bordone:2024hui}, described by the Lagrangian:
\bea
	\cL^{b\to s \llpair}_{\rm eff} = \frac{4G_F}{\sqrt{2}}  \frac{\alpha_e}{4\pi}\left(V^*_{ts}V_{tb}\sum_i C_i \cQ_i + \text{h.c.} \right)+ 	\cL^{N_f =5}_{\rm QCD\times  QED} \,,
	\label{eq:bsll}
\eea
The most relevant effective operators are: 
     \begin{align}
        \cQ_1 =& (\bar{s}^\alpha_{L}\gamma_\mu c^\beta_L)(\bar{c}^\beta_L\gamma^\mu b^\alpha_L)\,,   &  \cQ_2 =&(\bar{s}_{L}\gamma_\mu c_L)(\bar{c}_L\gamma^\mu b_L)\,,   \\
        \cQ_7=&\frac{e}{16\pi^2}m_b(\bar{s}_{L}\sigma^{\mu\nu} b_R)F_{\mu\nu}\,,    & \cQ_8=&\frac{g_s}{16\pi^2}m_b(\bar{s}_{L}\sigma^{\mu\nu}T^a b_R)G_{\mu\nu}^a\,,   \\
        \cQ_9=&\frac{e^2}{16\pi^2}(\bar{s}_{L}\gamma_\mu b_L)(\bar{\ell}\gamma^\mu \ell)\,,    &  \cQ_{10}=&\frac{e^2}{16\pi^2}(\bar{s}_{L}\gamma_\mu b_L)(\bar{\ell}\gamma^\mu\gamma_5 \ell)\,.
    \end{align}
To leading order in QED, the decay amplitude for $B \to M \llpair$, where $M=K, K^*$, can be written as:
\begin{equation}
    \begin{aligned}
&\cA (B \to M \llpair) = \frac{G_F \alpha_e V^*_{ts} V_{tb}}{\sqrt{2} \pi} \times \\
&\times \Big [(C_9 ~ \ell \gamma^\mu \ell + C_{10} ~ \ell \gamma^\mu \gamma_5 \ell ) \langle M | \bar{s} \gamma_\mu P_L b | \bar{B}\rangle  - \frac{1}{q^2} \ell \gamma^\mu \ell~ (2i m_b C_7 ~ \langle M|  \bar{s} \sigma_{\mu \nu} q^\nu P_R b | B\rangle + \mathcal{H}_\mu)  \Big],
    \end{aligned}
    \label{eq:nonlocal}
\end{equation}
where the first two matrix elements are local (determined by Lattice QCD and Light-Cone Sum Rules), and $\mathcal{H}_\mu$ contains the non-local contributions of the operators $\cQ_{1-6, 8}$. We parametrize the LD effects using dispersive relations to fit the data on $B \to K^{(*)}\mmpair$.
Finally, in our last approach \cite{Isidori:2024lng}, we estimate a class of LD effects that might mimic a New Physics (NP) contribution.  

\section{Comparison between experimental semi-inclusive rate and inclusive prediction}

Non-perturbative uncertainties in the high-$q^2$ region can be reduced by computing the ratio of the FCNC transition and the
$b\to u$ decay,
\be
 R_{\rm incl}(q_0^2) = \displaystyle\int_{q^2_0}^{m_B^2} d q^2 \frac{d \Gamma(B \to X_s \llpair)}{d q^2}  \Bigg / \displaystyle\int_{q^2_0}^{m_B^2} d q^2 \frac{d \Gamma(B \to X_u \bar\ell\nu)}{d q^2} \,,
 \label{eq:R0}
\ee
where $q_0^2 = 15 \text{~GeV}^2$. The hadronic structure of the two transitions is similar
($b\to q_{\rm light}$ left-handed current), leading to a significant cancellation of non-perturbative uncertainties. Thanks to the measurement of 
$\Gamma(B \to X_u \bar\ell \nu)$ by \cite{Belle:2021ymg}, this can be used to calculate  $\Gamma(B \to X_s \llpair)$. One finds:
\be
 R_{\rm incl}(q_0^2) = \frac{ |V_{tb} V^*_{ts}|^2 }{ |V_{ub}|^2 } \left[ \frac{\alpha_e^2 C_L^2 }{16 \pi^2} + \Delta \cR_{[q_0^2]} \right].
 \label{eq:RinclCVCL}
\ee
\begin{figure}[H]
\begin{minipage}[t]{0.65\textwidth}\vspace{0pt}
Here we have performed a change of basis from the usual $\{\cQ_9$, $\cQ_{10}\}$ to $\{\cQ_V, \cQ_L \}$ defined by $\cQ_V = (\overline{s}_L\gamma_{\mu} b_L)(\overline{\ell} \gamma^{\mu}\ell)$ and $\cQ_L = (\overline{s}_L\gamma_{\mu} b_L)(\overline{\ell}_L \gamma^{\mu}\ell_L)$.
The first term in Eq.~(\ref{eq:RinclCVCL}) is the result obtained in the limit of purely left-handed interactions and 
identical hadronic distributions, while $\Delta \cR_{[q_0^2]}$ describes all the deviations from this limit and the non-perturbative effects estimated in \cite{Ligeti:2007sn}.
We find: 
\be 
\mathcal{B}(B \to X_s \bar{\ell} {\ell})^{\mathrm{SM}}_{[15]} = (4.10\pm 0.81) \times 10^{-7}.
\label{eq:BRincl}
\ee
The leading uncertainties are due to the experimental result for $\Gamma(B\to X_u \bar{\ell}\nu)$ and the CKM inputs. We could therefore expect a 
significant reduction of the uncertainty in (\ref{eq:BRincl})
in the near future.
\end{minipage}
\hfill%
\begin{minipage}[t]{0.33\textwidth}\vspace{0pt}
\centering
    \includegraphics[width=0.75\linewidth]{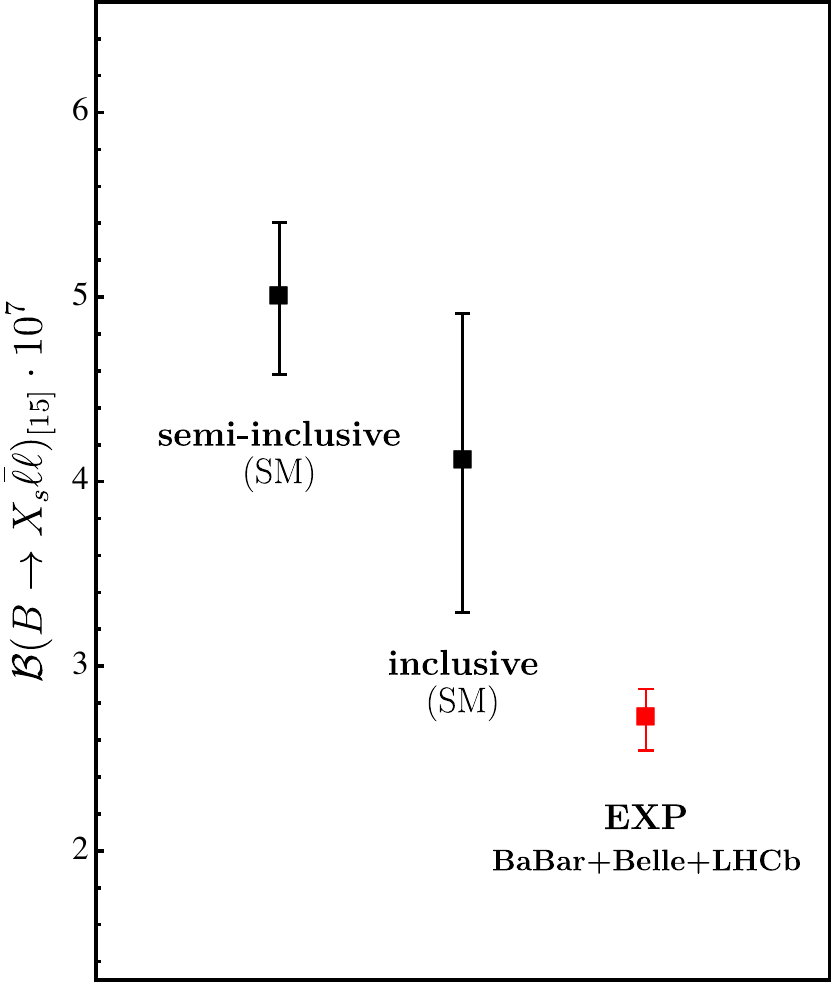}
    \caption{SM predictions vs.~experimental data for $\Gamma(B \to X_s \llpair)$.
    }
    \label{fig:inclusive}
\end{minipage} 
\end{figure}
On the experimental side, the $\Gamma(B \to X_s \mmpair)$ rate at high-$q^2$ is not fully available. However, in this kinematic region, only a few decay modes are relevant and we can replace the inclusive sum with the sum over a limited set of exclusive modes, namely $B \to K^* \mmpair$, $B \to K \mmpair$, $B \to K \pi \mmpair$, $B \to K \pi \pi \mmpair$, $B \to K \pi \pi \pi \mmpair$.
 We first show (Fig.~\ref{fig:inclusive}) the compatibility within the SM of the inclusive rate based on (\ref{eq:R0}) with a semi-inclusive calculation, where we include $B \to K^* \mmpair$, $B \to K \mmpair$, $B \to K \pi \mmpair$ (calculated in Heavy-Hadron Chiral Perturbation Theory (HHChPT)):
\bea
\sum_i \mathcal{B}(B \to X_s^i \bar{\ell} {\ell})^{\mathrm{SM}}_{[15]} = (5.07 \pm 0.42) \times 10^{-7}.
\eea
We then compare the inclusive prediction based on (\ref{eq:R0}) with an experimental sum-over-exclusive result based on \cite{LHCb:2014cxe}, \cite{LHCb:2016ykl}, \cite{LHCb:2014osj} including also BaBar and Belle data as calculated in \cite{Huber:2024rbw}:
\be
\mathcal{B}(B \to X_s \bar{\ell}{\ell})_{[15]}^{\mathrm{exp}} = (2.65 \pm 0.17)\times 10^{-7},
\ee
and find a $\approx 2 \sigma$ tension between these two values, in agreement with the tension on the exclusive modes.
\section{\texorpdfstring{Fit of $C_9$ from exclusive modes $B \to K^{(*)} \mmpair$}{}}

\begin{figure}[h!]
    \centering
    \includegraphics[scale=0.33]{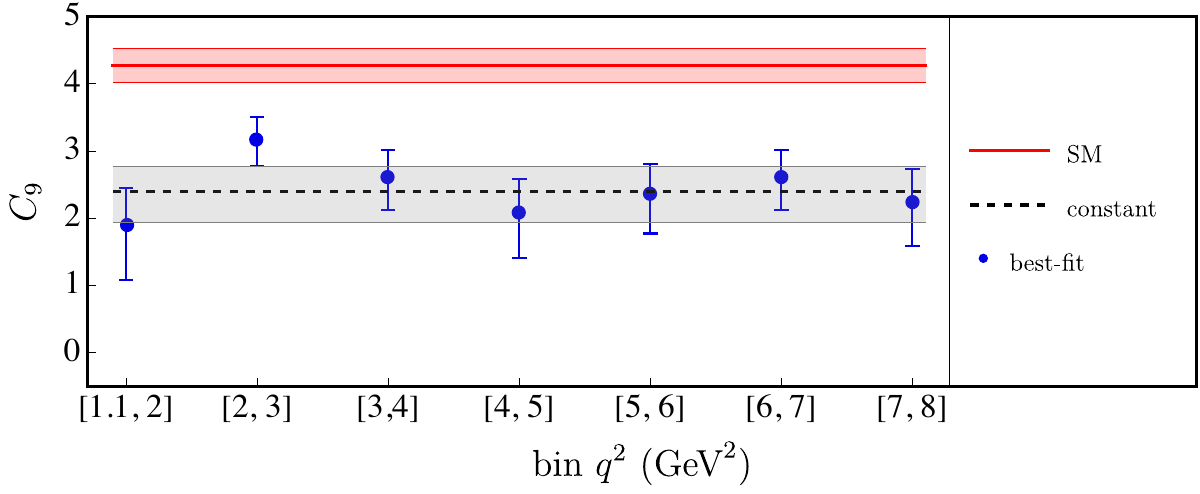}
    \includegraphics[scale=0.33]{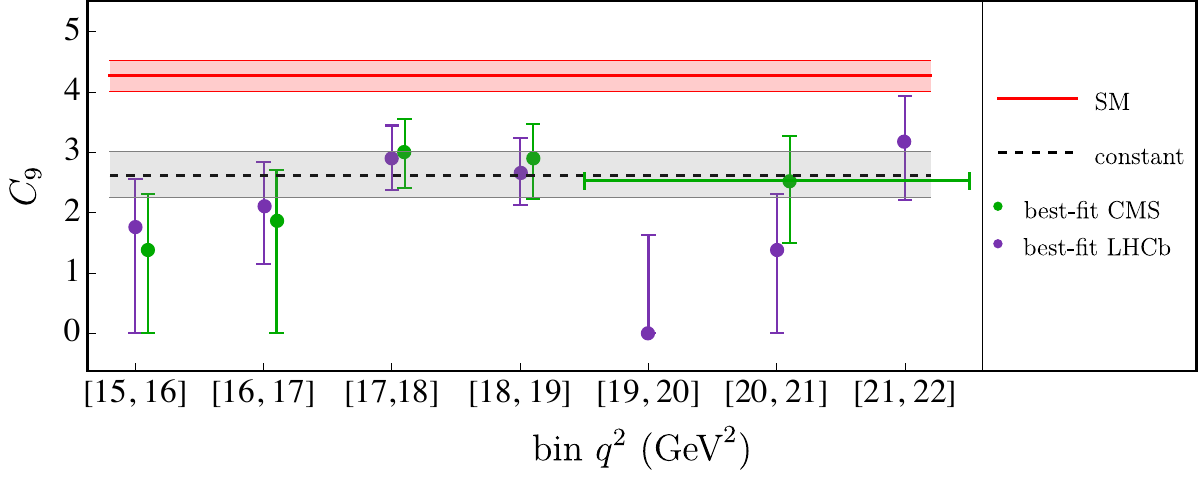}    

    \centering
    \includegraphics[scale=0.33]{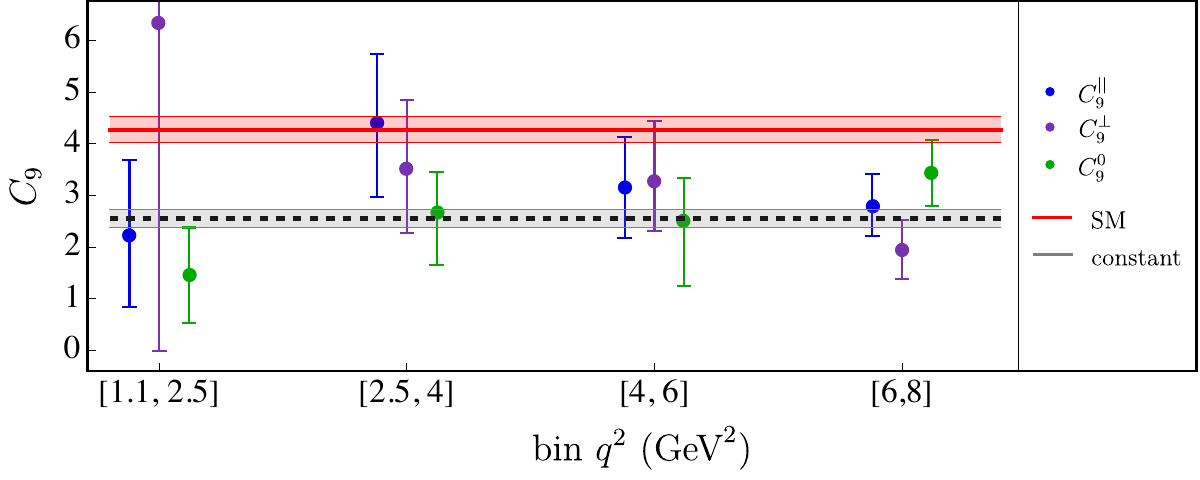}
    \includegraphics[scale=0.33]{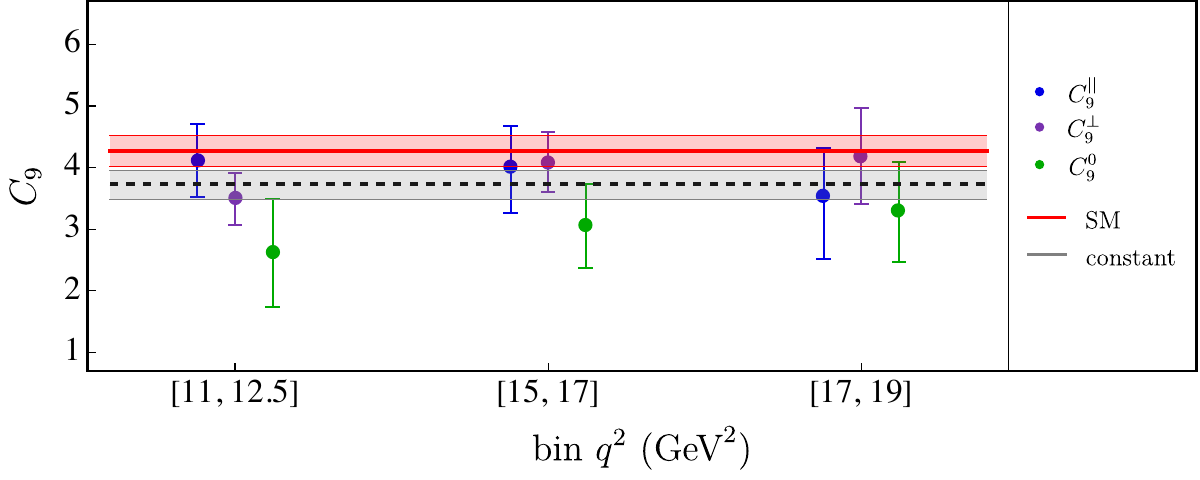}
    \caption{Determinations of $C_9$ in different $q^2$ bins  
    from $B \rightarrow K \mmpair$ (top) and $B \rightarrow K^* \mmpair$ (bottom). 
    The red and gray bands denote the SM value and the value extracted assuming 
    a $q^2$-independent $C_9$, respectively. }
        \label{fig:fit}
\end{figure}

As a further analysis, we look at the exclusive modes $B^+ \to K^+ \bar{\mu} \mu$ and $B^0 \to K^{*0} \bar{\mu} \mu$.
The non-local matrix elements in Eq.~\ref{eq:nonlocal}, to all orders in $\alpha_s$ and to first order in $\alpha_{em}$, have the same structure as the matrix elements of $\mathcal{O}_7$ and $\mathcal{O}_9$:
\begin{align}
 & \left.\cM\left(B\rightarrow M_\lambda \ell^+ \ell^-\right)\right\vert_{C_{1-6}}  =\ - i \frac{ 32 \pi^2\cN}{q^2}\,  \bar \ell \gamma^\mu \ell  
\int d^4 x e^{iqx} \langle H_\lambda   | T\left\{     
j_{\rm \mu}^{\rm em}(x), \sum_{i=1,6} C_i\cQ_i (0)
\right\} | B  \rangle \nonumber \\
& \qquad\qquad  = 
 \left( \Delta^\lambda_9(q^2) + \frac{m_B^2}{q^2} \Delta^\lambda_7
\right) \langle H_\lambda\ \ell^+ \ell^- |  \cQ_9  | B  \rangle \,.
\label{eq:MQi}
\end{align}
The (regular for $q^2 \to 0$) contributions of the non-local matrix elements of the four-quark operators can be effectively taken into account by a shift in $C_9$: $C_9 \to C_9^{\lambda}(q^2) + Y_{q \bar{q}}^{[0]}(q^2) +Y_{b \bar{b}}^{[0]}(q^2) + Y_{c\bar{c}}^\lambda (q^2)$, where the $Y$ functions include the factorizable perturbative contributions from four-quark operators, the perturbative charm loop contributions, and the $c\bar{c}$ resonances. To estimate the latter, we use dispersive relations in combination with data, where each resonance ($J/\psi$,  $\psi(2s)$,  $\psi(3770)$, $\psi(4040)$, $\psi(4160)$, and $\psi(4450)$) is parameterized by an amplitude and a phase, extracted from data \cite{LHCb:2016due}, \cite{LHCb:2024onj}.
We fit (Fig.~\ref{fig:fit}) from data the residual contribution $C_9^{\lambda}(q^2)$, which contains, other than the SM contribution, the unaccounted-for LD contributions, which would not be independent of $\lambda$ or $q^2$, and a possible SD new physics contribution, which instead would be independent of $\lambda$ and $q^2$.
For $B \to K \bar{\mu} \mu$ we perform a fit of $C_9$ bin by bin in $q^2$ by using the measured branching ratio by LHCb \cite{LHCb:2014cxe} and more recently by CMS \cite{CMS:2023klk}. In the low-$q^2$ region, since the experimental bins used by LHCb and by CMS are the same, we combine the two measurements.
For $B \to K^* \bar{\mu} \mu$ we perform the fit from the branching ratio and the angular observables measured by LHCb \cite{LHCb:2016ykl}.

\begin{figure}[h!]

\begin{minipage}[b]{.56\textwidth}
        \begin{tikzpicture}
                \begin{feynman}
                        \node (i1) {$B^0$};
                        \node[right=1cm of i1, dot] (i2);
                        \node[right=1.5cm of i2] (fake);
                        \node[above=0.866cm of fake, dot] (i3);
                        \node[below=0.866cm of fake, dot] (i4);
                        \node[right=1cm of i3] (j2) {$K^0$};
                        \node[right=1cm of i4] (j3) {$\gamma^*$};
                        \diagram*{
                                (i1) --[scalar] (i2) --[double] (i4) --[double] (i3) --[plain] (i2) ;
                                (i3) --[scalar] (j2) ;
                                (i4) --[photon] (j3) ;
                        };
                \end{feynman}
        \end{tikzpicture}
        \begin{tikzpicture}
                \begin{feynman}
                        \node (i1) {$B^0$};
                        \node[right=1cm of i1, dot] (i2);
                        \node[right=1.5cm of i2] (fake);
                        \node[above=0.866cm of fake, dot] (i3);
                        \node[below=0.866cm of fake, dot] (i4);
                        \node[right=1cm of i3] (j2) {$K^0$};
                        \node[right=1cm of i4] (j3) {$\gamma^*$};
                        \diagram*{
                                (i1) --[scalar] (i2) --[plain] (i4) --[plain] (i3) --[double] (i2) ;
                                (i3) --[scalar] (j2) ;
                                (i4) --[photon] (j3) ;
                        };
                \end{feynman}
        \end{tikzpicture}
        \caption{One-loop topologies considered in our analysis. Solid single lines denote charmed pseudoscalars ($D$ or $D_s$) and solid double lines denote charmed vectors ($D^*$ or $D^*_s$).}
        \label{fig:loops}
\end{minipage}
\begin{minipage}[b]{.34\textwidth}
\centering
  \includegraphics[scale=0.28]{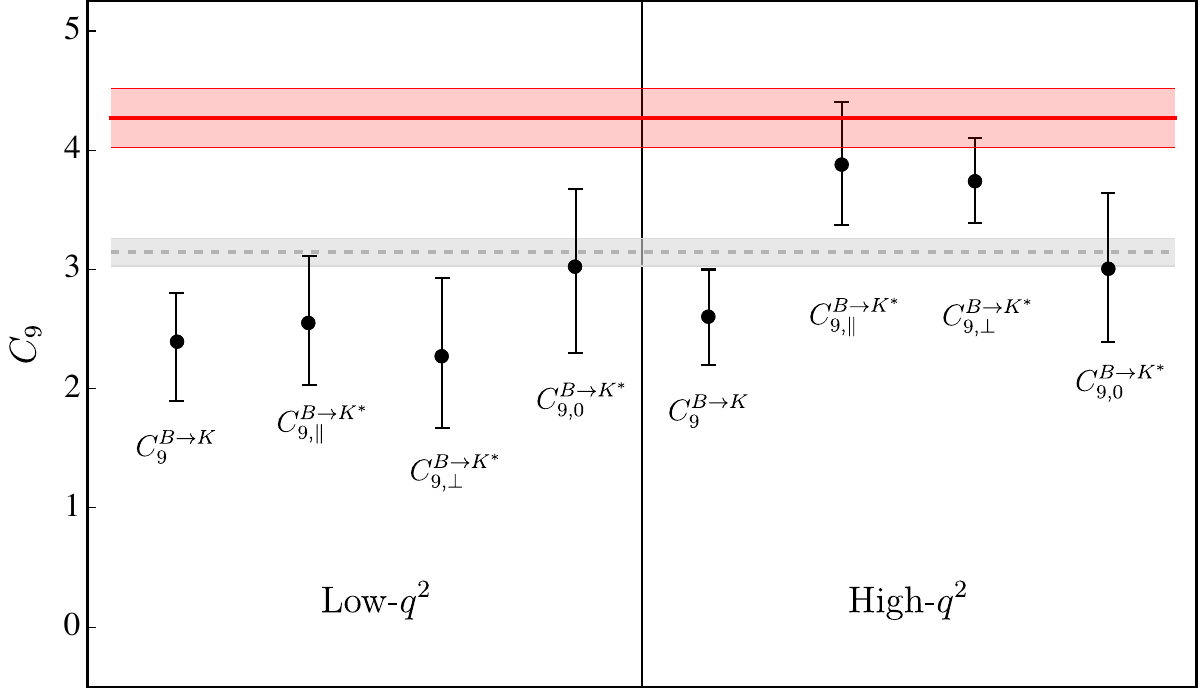}
    \caption{Independent determinations of $C_9$.}
        \label{fig:final}
\end{minipage}
\end{figure}

Except for the result at high-$q^2$ in $B \to K^* \mmpair$, we do not notice a significant $q^2$- or polarization-dependence, which would be present if we were missing the contribution of dominant LD effects.
In Fig.~\ref{fig:final} we show the best-fit results with the assumptions of $q^2$-independent $C_9$, in the low- and high-$q^2$ regions and for the different modes and polarizations.
 We also combine $B\to K$ and $B \to K^*$ in the full $q^2$ spectrum (gray dashed line).
These independent determinations of $C_9$ are compatible with each other, consistent with the hypothesis of the short-distance nature of the tension between the SM and data.

\section{\texorpdfstring{An explicit estimate of charm-rescattering effects in $B^0 \to K^0 \mmpair$}{}}
Lastly, we consider the LD effects in $B ^0 \to K^0 \llpair$ induced by the rescattering of a pair of charmed and charmed-strange mesons. We examine the leading contribution from the two-body intermediate state formed by a $ D D^*_s$ or $D^* D_s$ pair. Since we expect a reduced $q^2$-dependence by these types of contributions, these would mimic a SD effect, indistinguishable from NP.

Our model gives a description in terms of hadronic degrees of freedom supplemented by data on the $B \to D D^*$ decay. More precisely, the dynamics of the $D_{(s)}^{(*)}$ mesons close to their mass shell is imposed by the Lorentz transformation properties of the mesons, Gauge invariance under QED, $SU(3)$ light-flavor symmetry and heavy-quark spin symmetry. The $D D^* K$ and $D^{(*)} D^{(*)} \gamma$ vertices are estimated using HHChPT combined with the Vector Meson Dominance (VMD) Ansatz. With these tools, we obtain an accurate description in the low-recoil (or high-$q^2$) limit. To extrapolate our result over the entire kinematical range, we apply form factors to correct the $D^{(*)} D^{(*)} \gamma$ and the $ DD^*K$ vertices. For the former, we introduce an appropriate electromagnetic form factor to correct the point-like QED vertex, via a VMD Ansatz: we saturate the tower of narrow charmonium states by the leading $J/\Psi$ state.
For the $ DD^*K$ vertex, we notice that the model predicts a $K-$emission amplitude that grows like $E_K/f_K$ if $E_K$ is the kaon energy. This behavior needs to be corrected for $E_K > f_K$. Furthermore, the $1/f_K$ behavior of the amplitude can appear only for kaon momenta of $\mathcal{O}(\Lambda_{QCD})$. We therefore make the replacement: 
\begin{align}
\label{eq:kaonFF}
    \frac{1}{f_K} &\to \frac{1}{f_K} G_K(q^2)\, , ~~~~~ 
    G_K(q^2) = \frac{1}{1 + E_K(q^2)/f_K}
    = \frac{2 m_B f_K}{2m_B f_K + m_B^2 - q^2}\,.
    \nonumber
\end{align}
We compute the one-loop diagrams with internal $D^{(*)}$ and $D_s^{(*)}$ mesons with the model described; at the leading order, this amounts to the topologies depicted in Fig.~\ref{fig:loops}. In the $SU(3)$-symmetric limit, diagrams obtained by replacing $D \leftrightarrow D_s$ and $D^*_s \leftrightarrow D^*$ are identical.

\begin{figure}[!htb]
            \centering
        \includegraphics[width=0.35\linewidth]{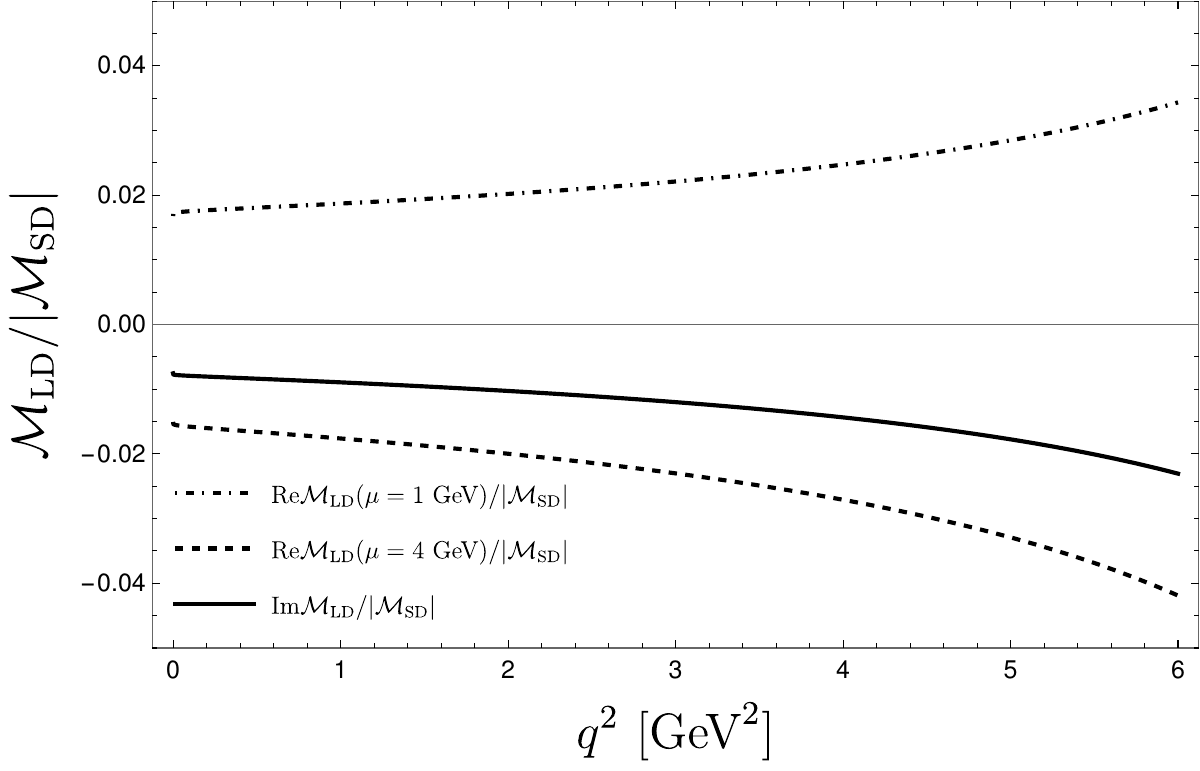}
         \includegraphics[width=0.35\linewidth]{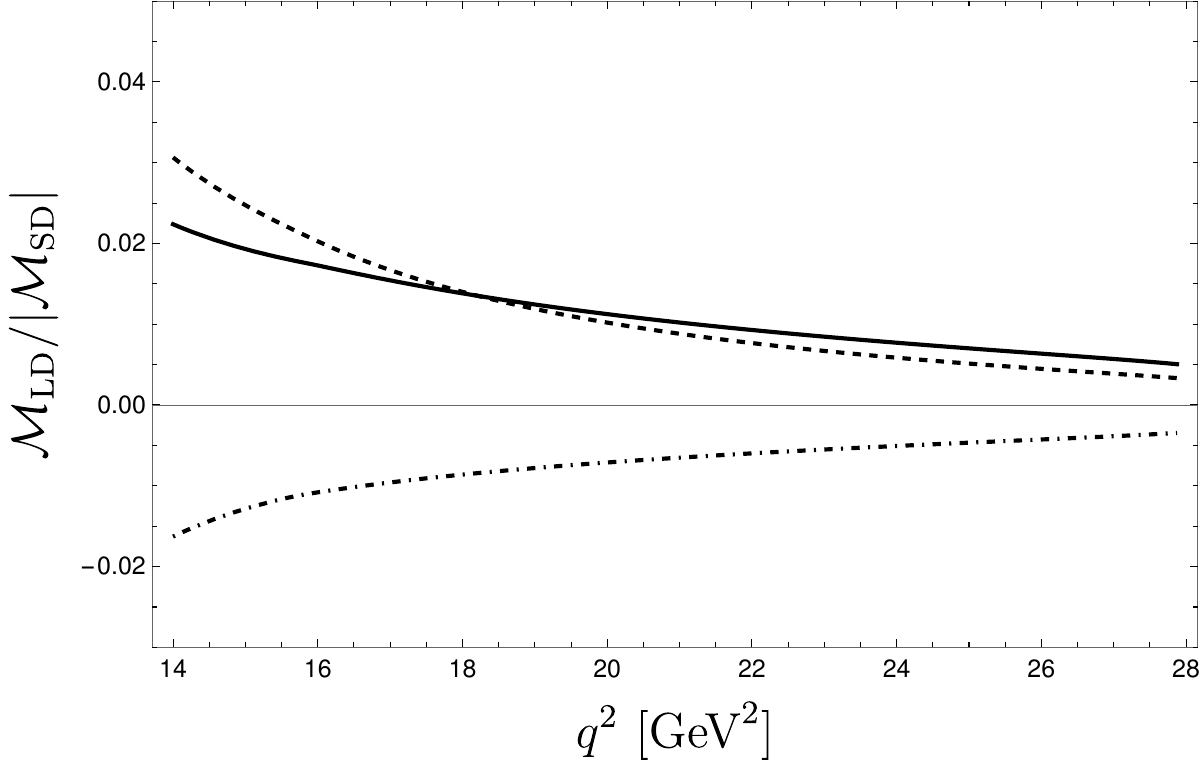}
        \caption{Ratio between charm-rescattering contributions
    to the matrix element without considering rescattering effects in the low-$q^2$ (top) and the high-$q^2$ (bottom) regions.}
          \label{fig:prob1_6_2}
\end{figure}

 The ratios of the matrix elements of these contributions over the SD contributions are shown in Fig.~\ref{fig:prob1_6_2}, where the dispersive and absorptive parts of the amplitude are plotted separately in the low- and high-$q^2$ regions. 
The sum of the diagrams has a UV-divergence that we discard using a $\overline{MS}$ scheme, thus introducing a renormalization scale dependence that we use to estimate the associated uncertainty. In particular, the absorptive part of the amplitude is independent of the renormalization scheme used. We can see that these LD contributions are relatively flat in $q^2$, and not large: encoding their effect in a shift in $C_9$, this is found to be of the order of $2.5 \%$ with respect to the SM value for the intermediate states considered.

Interestingly, the shift in $C_9$ has an opposite sign in the low- vs the high-$q^2$ region, hence comparing the values of $C_9$ extracted at different values of $q^2$ from data provides a useful data-driven check of the size of these LD contributions.

If we take into account all possible intermediate states that yield a parity-conserving strong interaction with the kaon, and we assume each gauge-invariant subset of diagrams adds coherently and roughly scales with the size of the corresponding $B^0 \to X_{\bar{c} c \bar{s} d}$ amplitude, we find a multiplicity factor of around 3. Therefore, these LD effects are encoded by a shift in $C_9$ of the order of $8-10 \%$ with respect to the SM. It is important to note that we neglected some effects, such as $SU(3)$-breaking effects ($\lesssim 30 \%$), higher-mass charmonium resonances ($N_c$-suppressed), baryonic modes ($\lesssim 10\%$), and higher-multipole photon couplings ($m_c$-suppressed), that are expected to be individually suppressed but could possibly lead to a larger-than-expected enhancement.

\end{document}